\documentclass[conference,a4paper]{IEEEtran}
\usepackage{cite}
\usepackage{bm}
\usepackage{multicol}
\usepackage{multirow}
\usepackage{amsmath} 
\usepackage{graphicx}
\usepackage{url}
\usepackage{color}
\usepackage{amsfonts}
\usepackage{amsmath}
\usepackage{extarrows}
\usepackage{graphicx}
\usepackage{amsfonts}
\usepackage{caption}
\usepackage{algorithm}
\usepackage{algpseudocode}
\usepackage{indentfirst}
\usepackage{caption}
\usepackage{esint} 
\usepackage{graphics}
\usepackage{epsfig}
\usepackage{soul}
\usepackage{caption}
\usepackage{gensymb}
\usepackage{algpseudocode}  

\usepackage{balance}
\usepackage{geometry}

\usepackage{algorithm,algpseudocode}
\makeatletter
\newcommand\fs@spaceruled{\def\@fs@cfont{\bfseries}\let\@fs@capt\floatc@ruled
  \def\@fs@pre{\vspace{1\baselineskip}\hrule height.8pt depth0pt \kern2pt}%
  \def\@fs@post{\kern2pt\hrule\relax}%
  \def\@fs@mid{\kern2pt\hrule\kern2pt}%
  \let\@fs@iftopcapt\iftrue}
\makeatother
\DeclareMathOperator{\trace}{tr}

\def\bA{{\mathbf{A}}}    
\def\bF{{\mathbf{F}}}  \def\bH{{\mathbf{H}}} \def\bI{{\mathbf{I}}} 
    
 \def\bQ{{\mathbf{Q}}}

\def\ba{{\mathbf{a}}}

  \def\br{{\mathbf{r}}}  \def\bt{{\mathbf{t}}}

\begin{document}
\title{Walsh Meets OAM in Holographic MIMO}
\author{Charles~Vanwynsberghe$^\dag$, Jiguang~He$^\dag$, Chongwen Huang$^\ddag$, and Merouane Debbah$^\dag$\\
        $^\dag$Technology Innovation Institute, 9639 Masdar City, Abu Dhabi, United Arab Emirates\\
        $^\ddag$College of Information Science and Electronic
Engineering, Zhejiang University, Hangzhou 310027, China
        }

\maketitle
\begin{abstract}
Holographic multiple-input multiple-output (MIMO) is deemed as a promising technique beyond massive MIMO, unleashing near-field communications, localization, and sensing in the next-generation wireless networks. Semi-continuous surface with densely packed elements brings new opportunities for increased spatial degrees of freedom (DoFs) and spectrum efficiency (SE) even in the line-of-sight (LoS) condition. In this paper, we analyze holographic MIMO performance with disk-shaped large intelligent surfaces (LISs) according to different precoding designs. Beyond the well-known technique of orbital angular momentum (OAM) of radio waves, we propose a new design based on polar Walsh functions. Furthermore, we characterize the performance gap between the proposed scheme and the optimal case with singular value decomposition (SVD) alongside perfect channel state information (CSI) as well as other benchmark schemes in terms of channel capacity. It is verified that the proposed scheme marginally underperforms the OAM-based approach, while offering potential perspectives for reducing implementation complexity and expenditure. 
\end{abstract}

\section{Introduction}
Holographic multiple-input multiple-output (MIMO) is considered as a cutting-edge technique beyond massive MIMO~\cite{Sha2018,Chongwen2020}. It unlocks the great potential of radiative near-field broadband communications by exploiting the numerous spatial degrees of freedom (DoFs) even under pure line-of-sight (LoS) condition. Thanks to densely packed antennas usually with less than half-wavelength inter-element spacing, highly directional pencil beams can be designed for ultra-high data rate transmissions.  

Due to the extreme densification of both transmit and receive elements, the channel state information (CSI) acquisition faces tremendous challenges. The reason lies in that the channel sparsity in the angular-domain representation vanishes when the receiver (Rx) moves from the far field to the near filed of the transmitter (Tx)~\cite{Mingyao2022}. In terms of precoding design, adopting the conventional right singular vectors of the channel (encounters high computational complexity), requires the availability of CSI information, and demands hardware flexibility to perform operations with variable amplitudes and phases. In the literature, the potential of communication modes and orbital orbital angular momentum (OAM) multiplexing was intensively investigated~\cite{Edfors2012,Dardari2020} with different antenna structures, e.g., uniform circular arrays (UCAs) and rectangular arrays. The OAM takes the advantage of electromagnetic (EM) wave property and generates orthogonal waves to multiplex various data streams at the transmitter to take full advantage of the available spatial multiplexing gain. Recently, two different types of OAM designs, i.e., focused and unfocused, were introduced in~\cite{Torcolacci2022}. Interestingly, no CSI is required\footnote{It should be noted that the Tx-Rx distance is needed to generate focused OAM modes.}, and continuous phase-shifters are sufficient to implement the precoder because the amplitude of OAM modes is constant.

Motivated by~\cite{Torcolacci2022}, we in this paper develop a design of communication modes based on polar Walsh functions. Such modes possess attractive properties: i) they are separable in radial and angular coordinates, ii) they are orthogonal, iii) their amplitude is constant, and their phases only take two values. The point iii) is interesting to substantially reduce the hardware complexity of the precoder and combiner\cite{Wang2018}. In addition, we compare the channel capacity with perfect CSI and the two benchmark schemes from~\cite{Torcolacci2022}. With a marginal channel capacity degradation with respect to the OAM modes \cite{Torcolacci2022}, the polar Walsh functions provide an interesting trade-off between communication performance and hardware complexity.

\textit{Notation:} A bold lowercase letter $\ba$ denotes the column vector, a bold capital letter $\bA$ denotes the matrix, $(\cdot)^{\mathsf{H}}$ and $(\cdot)^*$ denote the Hermitian transpose and conjugate, respectively, and $\|\cdot\|_2$ is the Euclidean norm.

\section{System Model}
\begin{figure}[t]
	\centering
    \includegraphics[width=0.95\linewidth]{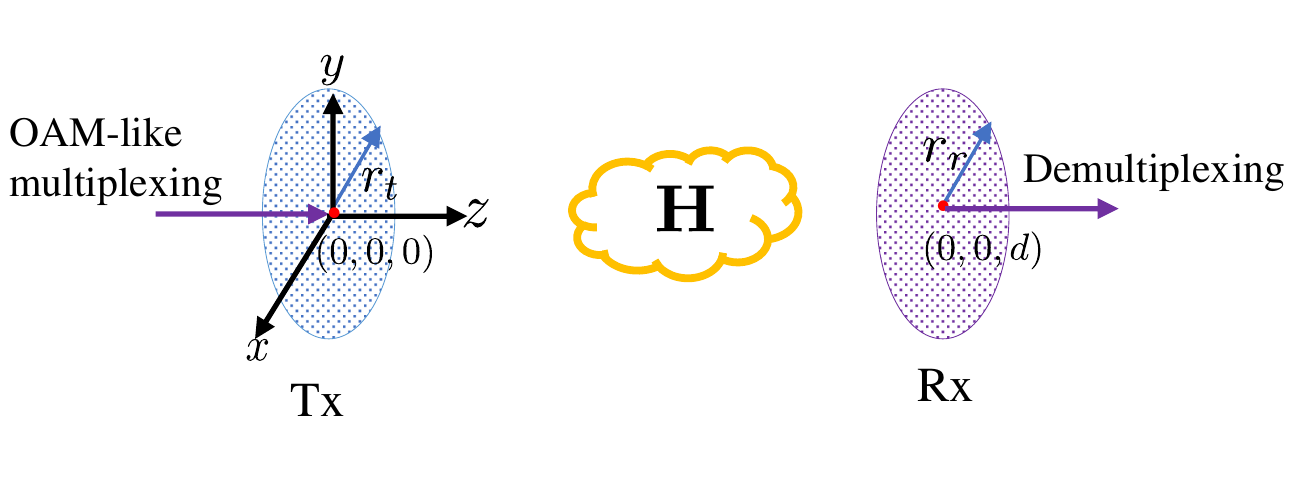}
	\caption{System model for OAM-like multiplexing based LIS communications.}
	\label{System_model}
\end{figure}

We leverage OAM-like multiplexing for data transmissions in the LIS-assisted communication network, shown in Fig.~\ref{System_model}, which is capable of processing the signals in the EM domain. The Tx and Rx are equipped with semi-continuous antenna apertures, centered at origin and point $(0,0,d)$ with areas $|\mathcal{S}_t| = \pi r^2_t$ and $|\mathcal{S}_r| = \pi r^2_r$, respectively, where $ \mathcal{S}_t=\{ (x,y, 0) \in \mathbb{R}^3: x^2 +y^2  \leq r_t^2 \} $, $ \mathcal{S}_r=\{ (x,y,d) \in \mathbb{R}^3: x^2 +y^2  \leq r_r^2 \}$, and $r_t$ and $r_r$ are the radii of the Tx and Rx apertures. These two antennas are thus parallel to the $x\text{-}y$ plane, centered onto the $z$ axis and separated by a distance $d$.

\subsection{Continuous Propagation Model}
The generated wave $\phi(\bt_0)$ at a certain Tx point $\bt_0 \in\mathcal{S}_t \subset \mathbb{R}^3$ and the received wave $\psi(\br)$ at a certain Rx point $\br\in\mathcal{S}_r \subset \mathbb{R}^3$ satisfy~\cite{Zidong2022,Torcolacci2022} 
\begin{equation}
    \psi(\br)  = \int_{\bt_0 \in \mathcal{S}_t} G(\br,\bt_0) \phi(\bt_0) \mathrm{d} \bt_0,
\end{equation}
where $G(\br,\bt_0)$ denotes the Green's function. Since we focus on near-field propagation regime, $G(\br,\bt_0)$ characterizes the wave propagation from the point $\bt_0$ to the point $\br$ in terms of phase change and attenuation. Therefore, $G(\br,\bt_0)$ is written in the form of 
\begin{equation}
\label{eq:greenfunc}
G(\br,\bt_0) = \frac{ \exp(-\jmath \kappa\|\br -\bt_0 \|_2)}{4 \pi \|\br -\bt_0 \|_2 },
\end{equation}
where $\kappa = 2\pi/\lambda$ with $\lambda$ being the wavelength and $\jmath = \sqrt{-1}$. 

Similar to singular value decomposition (SVD), we formulate a series of eigenproblems as 
\begin{equation}
    \label{eq:eigenproblem}
    \varepsilon \phi(\bt) = \int_{\bt_0 \in \mathcal{S}_t} K(\bt, \bt_0) \phi(\bt_0) \mathrm{d}\bt_0, 
\end{equation}
where the kernel $K(\bt, \bt_0)$ is defined as 
\begin{equation}
    K(\bt, \bt_0) = \int_{\br_0 \in \mathcal{S}_r} G^*(\br_0,\bt) G(\br_0,\bt_0) \mathrm{d} \br_0,
\end{equation}
and $\varepsilon$ is the eigenvalue associated with the eigenfunction $\phi(\cdot)$. Finding the solutions of this continuous eigenproblem is not straightforward. It is known from the literature that the circular prolate spheroidal wave functions (CPSWFs) are analytical solutions to the eigenproblem in Eq.~\eqref{eq:eigenproblem}. By resorting to CPSWFs, the DoFs of the channel between the transmitter and receiver can be calculated numerically~\cite{Zidong2022}.   

It is well-known that the optimal communication modes are provided by the solutions of the eigenproblem in Eq. \eqref{eq:eigenproblem}~\cite{telatar1999capacity}. Their numerical values can be obtained by discretizing the surfaces $\mathcal{S}_t$ and $\mathcal{S}_r$, reformulating the Eq. \eqref{eq:eigenproblem} under a matrix form and calculating the SVD~\cite{Dardari2020}. Although optimal, benefiting from these modes is challenging because it needs to handle different continuous values of amplitudes and phases in the precoder. Moreover, the exact knowledge of the CSI (i.e. Tx and Rx antenna rotations and relative distance) is required to obtain the eigenfuntions. The next section proposes alternative precoding schemes to tackle these limitations.

\section{Alternative designs of transmission modes}
\label{sec:tranmission_modes}

We present the different options of hardware-friendly transmission modes for the disk-shaped Tx LIS, with minimal requirements on the CSI. Following that logic, we introduce two families of  orthogonal modes:
\begin{itemize}
    \item OAM-based modes which depend on \textit{continuous} phases only with amplitude being \textit{constant};
    \item Polar Walsh functions which depend on \textit{binary} phases (i.e. $0$ and $\pi$) with amplitude being \textit{constant}. 
\end{itemize}
We define the polar coordinates on the Tx LIS as $(\rho, \theta) = (\sqrt{x^2 + y^2}, \arctan \frac{y}{x})$.

\subsection{OAM-based Transmission Modes}
In~\cite{Torcolacci2022}, two different types of OAM transmissions, i.e., focused and unfocused, are introduced and studied in terms of DoFs. The \textit{unfocused} OAM modes are defined for $n \in \mathbb{N}$ as:
\begin{equation}
\label{eq:oamu}
 \phi_n(\rho, \theta) =
 \begin{cases}
    \dfrac{1}{\sqrt{\pi}r_t} e^{- \jmath \frac{n-1}{2} \theta}, &\;\;\text{for $n$ even}, \\
   \dfrac{1}{\sqrt{\pi}r_t} e^{ \jmath \frac{n}{2} \theta},   &\;\;\text{for $n$ odd.}
  \end{cases}               
\end{equation}
This expression is a straightforward extension to the disk case of the OAM modes for UCAs \cite{Edfors2012}. Unfocused OAM modes are interesting because they do not need to know the distance $d$ between the Tx and Rx antennas.

The scenario in which $d$ is known is in favor of~\textit{focused} OAM modes, whose expression at the $n$-th state is:
\begin{equation}
\label{eq:oamf}
 \phi_n(\rho, \theta) =
 \begin{cases}
    \dfrac{1}{\sqrt{\pi}r_t} e^{\jmath \kappa d (1 + \frac{\rho^2}{2d^2})} e^{- \jmath \frac{n-1}{2} \theta}, &\;\;\text{for $n$ even}, \\
   \dfrac{1}{\sqrt{\pi}r_t} e^{\jmath \kappa d (1 + \frac{\rho^2}{2d^2})} e^{ \jmath \frac{n}{2} \theta},   &\;\;\text{for $n$ odd.}
  \end{cases}               
\end{equation}
Their expressions are based on an approximation of the exact CPSWFs that enables constant amplitude and preserves orthogonality~\cite{Jie2017}. To be specific, the phase varies also along the radial coordinate, so that the focal point is located at a distance $d$ from the Tx LIS. Both types of OAM modes are illustrated in Fig.~\ref{fig:modes} for $0 \leq n \leq 3$.

In this paper, we aim at augmenting the hardware-friendly simplification, and introduce a binary quantization of the modes. We will see that both radial and angular coordinates are separate functions of the original linear Walsh functions, and overall take only binary values $\pm \frac{1}{\sqrt{\pi}r_t}$. A cost-efficient and low-complexity implementation can be expected accordingly.

\subsection{Our Contribution: Transmission with Polar Walsh Functions}
\begin{figure}
    \centering
    \includegraphics[width=1\linewidth]{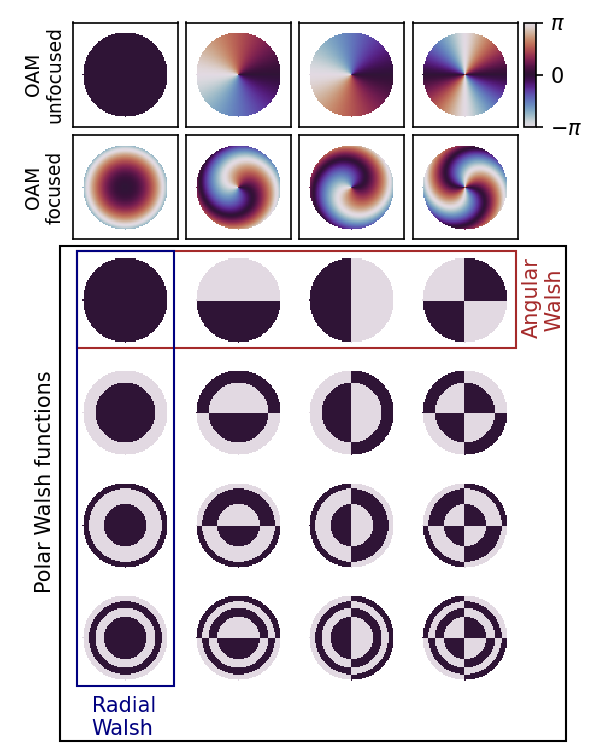}
    \caption{Phase plots of three orthogonal functions for holographic disk antennas: OAM unfocused, OAM focused ($r_t= 5\lambda$, $d =20\lambda$), and polar Walsh functions. Angular Walsh functions and radial Walsh functions are a subset of the complete polar Walsh functions.}
    \label{fig:modes}
\end{figure}

We present the Walsh functions in polar domain with separation in radial and angular coordinates \cite{Hazra2017}. The orthogonality of a series of polar Walsh functions is written as 
\begin{align}
    \int_0^{r_t} \int_0^{2\pi} \phi^{\mathcal{W}}_{m, n}(\rho, \theta) \phi^{\mathcal{W}}_{m', n'}(\rho, \theta) \rho\,\mathrm{d}\rho\,\mathrm{d}\theta &= \delta_{mm'} \delta_{nn'}, \\\nonumber 
    &\; \forall m, m', n', n',
\end{align}
where $\delta_{mm'}$ and $\delta_{nn'}$ are Kronecker delta functions. The index $m$ (resp $n$) of $\phi^{\mathcal{W}}_{m, n}$ stands for the communication mode state according to the radial (resp. angular) dimension. It can be shown from \cite{Hazra2017} that, for some positive integers $\mu$ and $\nu$, $\{\phi^{\mathcal{W}}_{m, n}\}$ forms an orthonormal basis for $0 \leq m \leq 2^\mu$ and $0 \leq n \leq 2^\nu$ with:
\begin{multline}
   \phi^{\mathcal{W}}_{m, n}(\rho, \theta)
   =
   \dfrac{1}{\sqrt{\pi} r_t} .
   \prod\limits_{k = 0}^{\mu-1} \mathrm{sgn}\bigg[\cos \Big(2^k \pi \left(\frac{\rho}{r_t}\right)^2 m_k\Big)\bigg] . \\
   \prod\limits_{k = 0}^{\nu-1} \mathrm{sgn}\bigg[\cos \Big(2^k \frac{\theta}{2} n_k\Big)\bigg] ,
\end{multline}
where $\mathrm{sgn}(\cdot)$ returns the sign of the argument, $m = \sum\limits_{k = 1}^{\mu-1} m_k 2^k$ with $m_k$ being the binary bits (0 or 1) of the binary numeral for $m$, ditto for $n = \sum\limits_{k = 1}^{\nu-1} n_k 2^k$. The polar Walsh function are illustrated in Fig. \ref{fig:modes} for $\mu = \nu = 2$.

Among the complete set of functions, the \textit{radial} Walsh functions are invariant with respect to  $\theta$, and generated by choosing $\nu=0$. Likewise, the \textit{angular} Walsh functions are invariant with respect to $\rho$, and generated by choosing $\mu=0$.

To the best of our knowledge, these three sets of functions are the only ones that (a) are defined on the disk, (b) are described with the minimal number of values -- two only -- and (c) form an orthonormal basis. Finally it is also independent from geometric parameters of the configuration. We emphasize that polar Walsh functions should not be strictly referred to as OAM modes, because they do not physically produce helical polarization of the EM waves.

\section{Capacity Formulation}
Literature related to OAM-based holographic MIMO focuses on numerical analysis of the DoFs. However, inter-mode interference may occur after propagation to the Rx when transmission modes are not the eigenfunctions from Eq. \eqref{eq:eigenproblem}. Here, we focus on the capacity metric with different transmission schemes.

Since the LIS aperture is continuous, we need to perform spatial sampling on the two-dimensional (2D) disk as to make the calculation of the channel capacity feasible. We assume that $N_t$ and $N_r$ points uniformly samples the Tx and Rx respectively, with a half-wavelength step size both $x$-wise and $y$-wise\footnote{It is mentioned in~\cite{Zidong2022} that half-wavelength sampling is sufficient for equivalently modeling the discretized antenna aperture
as its continuous counterpart without loss of accuracy by following Nyquist space sampling theorem.}. Thus, we can formulate the Tx-Rx propagation channel with a matrix $\bH \in \mathbb{C}^{N_r \times N_t}$, whose entries are calculated based on the positions of the Tx and Rx elements and the model Eq. \eqref{eq:greenfunc}. Therefore, the channel capacity $C$ in [bit/s/Hz] is written as 
\begin{equation}
C = \log_2 \det \bigg( \bI_{N_r} + \frac{P_t}{\sigma^2} \bH \bF \mathbf{Q} \bF^\mathsf{H} \bH^\mathsf{H} \bigg),
\end{equation}
where $\dfrac{P_t}{\sigma^2}$ denotes the received signal-to-noise ratio (SNR), the $N_t \times N$ matrix $\bF$ contains a collection of $N$ selected modes (similar to precoding in the conventional MIMO network), and $\bQ$ is the $N \times N$ symmetric matrix that allocates the power to each mode, such that $\trace(\bQ) \leq 1$. In order to analyze the capacity at constant received SNR, the total transmitted power is adjusted to compensate the path loss, as
\begin{equation}
 P_t = \text{SNR} \cdot  (4\pi d \sigma)^2.
\end{equation}

When the channel matrix $\bH$ is known, the capacity is maximized when $\bF$ consists of the right singular vectors from $\bH$, and when the power is distributed by water-filling \cite{telatar1999capacity}. In that specific case $\mathbf{Q}$ forms a diagonal matrix. Another sub-optimal yet reasonably efficient approach, named equal power allocation (EPA), distributes the power on each mode equally \cite[Appendix C]{Marzetta2016} with 
\begin{equation}
 \mathbf{Q} = \dfrac{\bI_N}{N}.
\end{equation}
EPA is interesting in the general case of \textit{any} choice of precoder $\bF$, including OAM modes and Walsh functions. Indeed, finding the matrix $\mathbf{Q}$ that maximizes $C$ under constrained transmission power deals with the power allocation problem, which goes beyond the scope of our study. First and foremost, we want to compare the possible candidates described in Sec. \ref{sec:tranmission_modes} to design the precoder $\bF$ independently from the power allocation subject. Therefore, the next section analyzes the capacity under EPA with different types of precoders $\bF$.
 
\section{Numerical Results}
In this section, we evaluate the channel capacity of the different transmission modes described in Sec. \ref{sec:tranmission_modes}. The configuration is set with $r_t = r_r = 10\lambda$, $\lambda = 1$~cm and the SNR equals $-20$~dB. For the study, the following transmission schemes are compared under EPA, such that $\bF$ contains:
\begin{itemize}
    \item the first $N=16$ right singular vectors of $\bH$;
    \item the first $N=16$ OAM unfocused modes ($0\leq n \leq 15$ in Eq. \eqref{eq:oamu});
    \item the first $N=16$ OAM focused modes ($0\leq n \leq 15$ in Eq. \eqref{eq:oamf});
    \item the first $N=16$ radial Walsh functions ($\mu = 4$, $\nu = 0$);
    \item the first $N=16$ angular Walsh functions ($\mu = 0$, $\nu = 4$);
    \item the first $N=16$ polar Walsh functions ($\mu = 2$, $\nu = 2$).
\end{itemize}
The comparison is also made with the optimal water-filling case.

\begin{figure}
    \centering
    \includegraphics[width=1\linewidth]{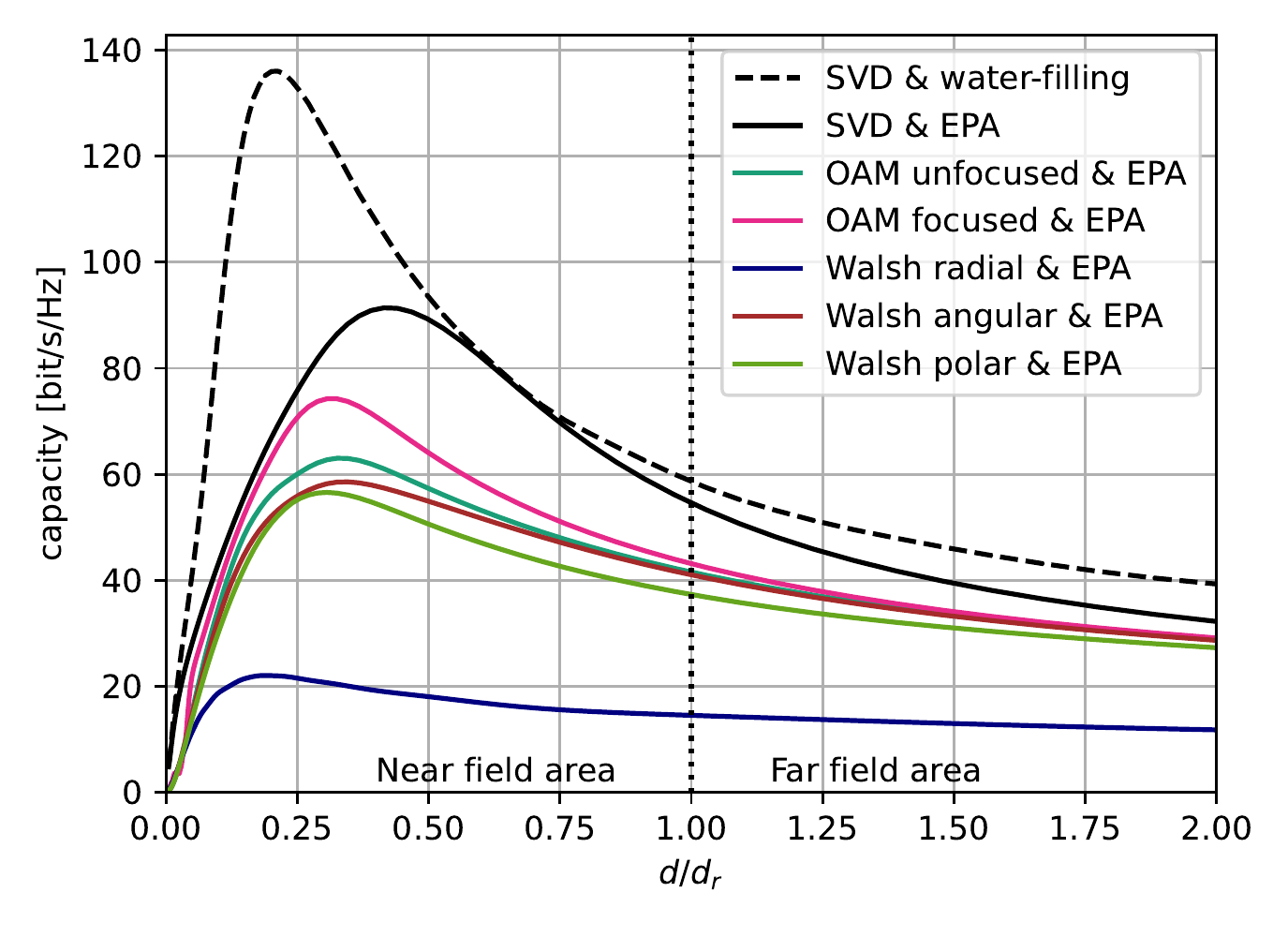}
    \caption{Channel capacity comparison with disk-shaped LISs. The $x$-axis is the ratio between the Tx-Rx distance and Rayleigh distance. }
    \label{fig:capacity}
\end{figure}
 
The capacity is plotted in Fig.~\ref{fig:capacity} as a function of distance $d$ scaled to the Rayleigh distance $d_r = \dfrac{r_t^2}{2 \lambda}$. The vertical dotted line discriminates between near field and far field regions. As expected, the SVD modes with water-filling dominate all the other candidate transmission schemes. The bell-shaped trend with a maximum in the near field (here $\approx 136$ bit/s/Hz) is a well known behavior of the continuous apertures when they are closely spaced \cite{Zidong2022}. With EPA, SVD still outperforms all the other schemes, since it benefits from the perfect CSI. 

Except for the radial Walsh function that substantially has the lowest performance, all the other precoding designs attain a capacity in a range going from $56$ to $74$ bit/s/Hz. The focused OAM functions bring a slight improvement over the unfocused ones thanks to the knowledge of $d$, particularly in the near field. However, polar and angular Walsh functions attain capacities at the same order of magnitude than OAM modes from~\cite{Torcolacci2022}, with a marginal degradation in the near-field regime.

Interestingly, going from complex-valued (OAM) to binary-valued (Walsh) modes has little impact on the communication performance, whereas it potentially leads to significant hardware simplifications of the Tx precoder \cite{Wang2018}. Note that these trends are given for $N = 16$, and choosing other small values (e.g. $N<25$) does not compromise the stated conclusions. As potential  extension of the current study, the authors suggest to investigate a power allocation method with the given the precoding constraint to further improve the capacity. Investigating the actual benefits to the hardware complexity is also of importance, since the required element densification in holographic MIMO may lead to very large values of $N_t$.

\section{Conclusion}
In this paper, we have proposed a cost-efficient and hardware-friendly holographic MIMO transmission scheme for disk apertures based on polar Walsh functions. Rather than focusing on DoFs, we opt for a numerical analysis of the capacity. Even though the proposed scheme exhibited slight performance degradation compared to the state-of-the-art OAM schemes, our proposed approach is more readily to be implemented/realized in the future LIS-assisted network. In the future, we will focus on the post-processing of the received signals and conduct rigorous transmit power allocation in order to achieve the full potential of OAM based LIS communications.  
  
 \balance
\bibliographystyle{IEEEtran}
\bibliography{IEEEabrv,Ref}

\end{document}